\documentclass[pra,aps,showkeys,twocolumn,10pt,amsmath,amssymb]{revtex4-2}
\pdfoutput=1

\usepackage{bm}
\usepackage[utf8]{inputenc}
\usepackage[T1]{fontenc}
\usepackage[dvipsnames]{xcolor}
\usepackage{braket}
\usepackage{amsthm}
\usepackage{tikz}
\usepackage{caption,subcaption}
\usepackage{tikz-network}

\newcommand{\Idist}{\ensuremath{\hat{I}_{\mathrm{dist}}}}
\newcommand{\mc}{\mathcal}
\newcommand{\bs}{\boldsymbol}
\DeclareMathOperator{\tr}{Tr}
\DeclareMathOperator*{\rank}{rank}
\DeclareMathOperator*{\range}{range}
\newcommand{\Umin}{\ensuremath{U_{\mathrm{min}}}}

\newtheorem*{theorem}{Theorem}

\begin{document}

\title{On the Ordering of Sites in the Density Matrix Renormalization Group\\
using Quantum Mutual Information}

\author{Mazen Ali}
\email{mazen.ali@ec-nantes.fr}
\affiliation{Laboratoire de Mathématiques Jean Leray\\
Centrale Nantes, France}

\date{\today}

\begin{abstract}
    The density matrix renormalization group (DMRG) of White 1992 remains to this
    day an integral component of many state-of-the-art methods for
    efficiently simulating strongly correlated quantum systems.
    In quantum chemistry, QC-DMRG became a powerful tool for
    \emph{ab initio} calculations with the non-relativistic Schrödinger equation.
    An important issue in QC-DMRG is the so-called \emph{ordering problem}
    -- the optimal ordering of DMRG sites corresponding
    to electronic orbitals that produces the most accurate results.
    To this end, a commonly used heuristic is the grouping of strongly
    correlated orbitals as measured via quantum mutual information (QMI).
    In this work, we show how such heuristics can be directly related to minimizing
    the entanglement entropy of matrix product states (MPS)
    and, consequently, to the truncation error of a fixed bond dimension approximation.
    Key to establishing this link is the strong subadditivity of
    entropy. This provides rigorous theoretical justification for the orbital ordering methods
    and suggests alternate ordering criteria.
\end{abstract}

\keywords{Density Matrix Renormalization Group (DMRG), Matrix Product States (MPS),
Tensor Networks, Orbital Ordering,
Electronic Schrödinger, Entanglement Entropy, Quantum Mutual Information}

\maketitle

\section{Introduction}

Although quantum many-body systems are a priori described by the Schrödinger equation,
its accurate solution is a notoriously difficult problem. Early renormalization group
ideas attempted to address this issue by focusing on re-scaling transformations
between energy or length scales that, ideally, in each step reduce the number of degrees
of freedom while maintaining a good approximation \cite{Wilson75}.
In \cite{White92}, White proposed the density matrix renormalization group (DMRG)
-- a breakthrough numerical algorithm that allowed for highly accurate
computations of 1D quantum lattice systems. An important insight
from \cite{White92,Schollwock2011} is that, in each renormalization step,
an accurate approximation is achieved by retaining the degrees of freedom
necessary for an accurate description of the entanglement structure.
White achieved this by retaining the principal eigenvectors of the reduced density
matrix corresponding to the largest eigenvalues -- giving the method its name.
To this day DMRG remains a crucial component of modern methods for
quantum many-body problems.

In \cite{White99}, DMRG was extended to applications in quantum chemistry (QC-DMRG)
and has witnessed remarkable success ever since, see
\cite{Wouters2014, Szalay2015} for a review.
In QC, within the Born-Oppenheimer approximation, using the full
configuration interaction (FCI) ansatz and second
quantization, one can represent the wave function
in terms of the occupied orbitals together with a one-orbital
basis set, e.g., $\set{\ket{0}, \ket{\uparrow}, \ket{\downarrow},
\ket{\uparrow\downarrow}}$. After this DMRG can be applied
within the occupation representation.

The efficiency of DMRG relies on an accurate representation of the entanglement
structure which in turn depends on the ordering of \emph{sites}.
In QC-DMRG, a site corresponds to an orbital.
The correlation between different orbitals depends on
the choice of the single-particle basis and the ordering of the orbitals
on a 1D chain.
Unlike for a 1D lattice system,
a good ordering of orbitals in QC applications with non-local
interactions is not a priori clear.
It is by now well-known that optimizing said ordering leads to
substantial gains in accuracy
\cite{Legeza2003,Moritz2005,Rissler2006,Ghosh2008,Murg2010,Marti2010,Barcza2011,
Mitrushchenkov2012,Keller2014,Szalay2015,Murg2015,Krumnow2016,Dupuy21}.
The correlation
structure within DMRG is affected by the choice of the single-particle
basis as well, see, e.g., \cite{Szalay2015,White2018}.

There have been several approaches to ordering optimization.
In this work, we discuss one of the most successful ones, introduced in
\cite{Rissler2006,Barcza2011}.  It relies on an analysis of the \emph{quantum mutual information}
(QMI)
$I_{i,j}$ between orbitals $i,j$.
The authors propose to order the orbitals
such that the entanglement distance\footnote{In \cite{Rissler2006},
the authors minimize $\Idist^-$ with $\eta=-2$.
In \cite{Barcza2011}, the authors minimize $\Idist^+$
with $\eta=1,2$.}
\begin{align}\label{eq:Idist}
    \Idist^\pm:=\pm\sum_{i,j}I_{i,j}\times|i-j|^\eta,
\end{align}
is minimized.
This forces strongly correlated orbitals to be grouped
closer together on the chain.
To avoid the inevitable combinatorial complexity of testing
all possible orbital configurations, the authors propose using an approximate
optimization scheme.
The latter point is not relevant
to this discussion as we focus on $\Idist^\pm$ as a measure of entanglement and
approximability.
We also note the recent ordering criteria proposed in \cite{Dupuy21} which
outperforms \cite{Barcza2011} in several numerical experiments.
We comment on this criteria in Appendix \ref{app:misong}.

Although minimizing $\Idist^\pm$ is based on sound entanglement principles,
it does not fully explain why optimizing this criteria has such a significant
effect on the approximation accuracy of DMRG.
In this work, we demonstrate that the link between two-orbital
entanglement and DMRG approximation accuracy can be made quite
rigorous.
DMRG can be seen as a variational ansatz over a matrix product state
(MPS) \cite{Ostlund95}. It is known \cite{Schuch08,Cirac06}
that the approximation accuracy
of a fixed bond MPS is intimately related to the entanglement entropy
of subchains. The latter, as we will demonstrate, is bounded by the two-site
QMI $I_{i,j}$. This holds for any entropy measure
satisfying the subadditivity (SA) property. We thus provide a rigorous justification
for using $\Idist^\pm$ as a measure of accuracy and entanglement of 1D chains.
This also suggests alternate minimization criteria that are more precise
bounds for the subchain entropy, see \eqref{eq:Imps}.
We discuss the role of different SA properties and entanglement measures.

We remark that in this work we loosely refer to
\emph{entanglement} even though QMI measures total
correlation which includes classical correlation as well.
In a recent work \cite{Ding2021} this was addressed in detail. This does
not affect the issue of estimating block entropy.

In Sections \ref{sec:mps} and \ref{sec:ent},
we briefly review MPS, entanglement entropy and the link to
approximation accuracy. Section \ref{sec:qmi} contains the main technical result.
We conclude in Section \ref{sec:discuss}
with a discussion about $\Idist^\pm$, alternate criteria and
different entropy measures.
In the Appendix, we provide a proof of the main result \eqref{eq:main},
a brief discussion on the
recent ordering criteria from \cite{Dupuy21}, tree tensor networks and
point out a connection between different entropy
measures and notions of low-rank approximability.
\section{Matrix Product States}\label{sec:mps}

As was observed in \cite{Ostlund95}, DMRG converges to a fixed point that can be written
down as an MPS
\begin{align}\label{eq:mps}
    \ket{\Psi}=\sum_{s_1,\ldots,s_L}^d U^{s_1}\cdots U^{s_L}
    \ket{s_1\cdots s_L},
\end{align}
where $U^{s_k}$ are matrices of size $\chi_{k-1}\times\chi_k$,
the numbers $\chi_k$ are referred to as \emph{bond dimensions},
and $d$ is the size of the local Hilbert space.
For the purposes of this work we only consider finite size MPS with
open boundary conditions $\chi_0=\chi_L=1$.
The contraction of the matrices $U^{s_k}$
is represented pictorially in a tensor diagram in Figure
\ref{fig:tensors} as a 1D chain.
Such representations can be easily generalized to more complex
networks and so an MPS is a particular kind of a tensor network \cite{Orus2019}.

In QC-DMRG, in second quantized form, the basis is given as
\begin{align*}
    \ket{s_1\cdots s_L} = (a_1^\dagger)^{s_1}\cdots(a_L^\dagger)^{s_L}
    \ket{0},
\end{align*}
where $a_k^\dagger$ is the creation operator for the $k$-th orbital.
The basis thus represents occupation numbers of the corresponding orbitals,
e.g., $\ket{s_k}\in\set{\ket{0}, \ket{\uparrow}, \ket{\downarrow},
\ket{\uparrow\downarrow}}$,
and in this case $d=4$.

An MPS can represent any state in the full tensor product Hilbert space
provided the bond dimensions are \emph{sufficiently large}.
For a generic state, the bond dimension will grow exponentially
in the system size. But it is by now well understood that for many
physical states of interest -- such as low-lying states of 1D Hamiltonians
-- a fixed bond MPS provides an accurate approximation \cite{Eisert2010}.
This provides theoretical justification for using DMRG.

The truncation error incurred for a fixed bond dimension MPS can be analyzed
via the \emph{Schmidt decomposition}. Any pure state of a bipartite system
can be decomposed as
\begin{align*}
    \ket{\Psi_{AB}}=\sum_k \sigma_k^{[A]}\ket{k_A}\otimes\ket{k_B},
\end{align*}
with $\sigma_1^{[A]}\geq\sigma_2^{[A]}\geq\ldots\geq 0$.
Retaining only the first $\chi$ states, the resulting truncation error is
\begin{align}\label{eq:errorA}
    \|\ket{\Psi_{AB}}-\ket{\Psi_{AB}^{\mathrm{trunc}}}\|^2=
    \sum_{k>\chi}(\sigma_k^{[A]})^2.
\end{align}

For an MPS such a truncation can be applied successively to different subchains
\cite{Vidal2004,Oseledets2011,Vidal2003,Grasedyck2010}.
In each step, $A:=\set{1,\ldots,j}$ and
$B:=\set{1,\ldots,L}\setminus A$ and we denote by
$\varepsilon^2_j(\chi)$ the truncation error from
\eqref{eq:errorA} for bond dimension $\chi$.
The total truncation error is subadditive
\begin{align}\label{eq:trunc}
    \|\ket{\Psi}-\ket{\Psi^{\mathrm{trunc}}}\|^2&\leq
    \sum_{j=1}^{L-1}\varepsilon_j^2(\chi)=
    \sum_{j=1}^{L-1}\sum_{k>\chi}(\sigma_k^{[1,\ldots,j]})^2\notag\\
    &=:\varepsilon^2(\chi).
\end{align}
The efficiency of DMRG relies on $\varepsilon^2(\chi)$ remaining small for
not too large $\chi$. Note that, strictly speaking, \eqref{eq:trunc} is an idealized situation:
DMRG does not rigorously guarantee convergence to a global
minimum \cite{Chan2002,Holtz2012}, and so, in that sense, \eqref{eq:trunc} is the best error one
can hope for. Nonetheless, in practice
-- applying possibly modifications to DMRG \cite{Moritz06,Dolgov2014} -- $\varepsilon^2(\chi)$
provides a good a priori estimate of its performance.
\section{Entanglement Entropy}\label{sec:ent}

The question of simulability with MPS is closely related to the deeper notion of
entanglement entropy.
A contiguous block $A$ as in Figure \ref{fig:tensors}
may be in general entangled
with its environment. The reduced state of $A$ is then described by a
\emph{density operator} $\rho^{[A]}:\bs{\mc H}_A\rightarrow\bs{\mc H}_A$
\begin{align*}
    \rho^{[A]}=\sum_k\lambda_k^{[A]}\ket{k_A}\bra{k_A},
\end{align*}
with ordered probabilities $\lambda_1^{[A]}\geq\lambda_2^{[A]}\geq\ldots\geq 0$
summing to $1$.

To quantify entanglement, we will use a measure of the entanglement
entropy. The most commonly used measure is the
\emph{von Neumann entropy}
\begin{align*}
    S(\rho^{[A]}):=-\tr(\rho^{[A]}\log_2 \rho^{[A]})=
    -\sum_k\lambda_k^{[A]}\log_2(\lambda_k^{[A]}).
\end{align*}
It is the quantum analogue of the Shannon entropy,
and it quantifies the amount of uncertainty about the
state $\rho^{[A]}$. Another common measure is the
\emph{R\'{e}nyi entropy}
\begin{align*}
    S^\alpha(\rho^{[A]})&:=(1-\alpha)^{-1}\log_2\tr([\rho^{[A]}]^\alpha)\\
    &=(1-\alpha)^{-1}\log_2\sum_{k}(\lambda_k^{[A]})^\alpha,
\end{align*}
with $\alpha>0,\;\alpha\neq 1$.
For $\alpha\searrow 1$, $S^\alpha(\rho^{[A]})\rightarrow S(\rho^{[A]})$
and so one can denote the von Neumann entropy as
$S^{\alpha=1}(\rho^{[A]})$.
Another interesting limit is the
\emph{Hartley entropy}
$S^{\alpha=0}(\rho^{[A]})=\lim_{\bar\alpha\searrow 0}
S^{\bar\alpha}(\rho^{[A]})=\log_2\rank(\rho^{[A]})$,
see also Appendix \ref{sec:apphart}.

For the MPS from \eqref{eq:mps}, applying the Schmidt decomposition
for the bipartite splitting into subchain
$A:=\set{1,\ldots,j}$ and environment $B:=\set{j+1,\ldots,L}$ yields
\begin{align*}
    \ket{\Psi}=\sum_{k}\sigma_k^{[1,\ldots,j]}\ket{k_{1,\ldots,j}}\otimes
    \ket{k_{j+1,\ldots,L}}.
\end{align*}
This gives the reduced density matrix
\begin{align*}
    \rho^{[1,\ldots,j]}&:=\tr_{\set{j+1,\ldots,L}}(\ket{\Psi}\bra{\Psi})\\
    &=\sum_k(\sigma_k^{[1,\ldots,j]})^2\ket{k_{1,\ldots,j}}\bra{k_{1,\ldots,j}}.
\end{align*}
We measure the entanglement entropy of this subchain as
\begin{align*}
    S^\alpha_{[1,\ldots,j]}:=S^\alpha(\rho^{[1,\ldots,j]}),\quad\alpha\geq 0.
\end{align*}

If the entanglement entropy $S$ scales proportionally to
the area $|\partial A|$, then such systems are said
to satisfy an \emph{area law} \cite{Schuch08,Eisert2010}.
Area laws in general do not guarantee approximability
with tensor networks \cite{Ge2016}.
But for the special case $D=1$,
an area law (see Figure
\ref{fig:mps}) implies $S\sim 1$
and thereby approximability \cite{Cirac06}.

As shown in \cite{Cirac06,Schuch08}, for \emph{any} state, one can bound the truncation error
$\varepsilon_j(\chi)$ by the entanglement entropy of the corresponding
subchain
\begin{alignat}{2}\label{eq:simbound}
    \varepsilon_j(\chi)&\leq \left(\frac{\chi}{1-\alpha}\right)^{(\alpha-1)/\alpha}
    2^{\frac{1-\alpha}{\alpha}S^\alpha_{[1,\ldots,j]}}, &&\quad0<\alpha<1,\\
    \varepsilon_j(\chi)&\geq 1-\chi^{(\alpha-1)/\alpha}
    2^{-\frac{\alpha-1}{\alpha}S^\alpha_{[1,\ldots,j]}},&&\quad\alpha>1.\notag
\end{alignat}
Note that R\'{e}nyi entropies are monotonically descreasing \cite{Renyi1961}: $S^{\alpha_1}\geq S^{\alpha_2}$ for $0\leq \alpha_1\leq\alpha_2$. An upper bound for
$\varepsilon_j(\chi)$ trivially holds using the Hartley entropy $S^{\alpha=0}$
but this is not useful for approximation.

The von Neumann entropy $S^{\alpha=1}$ represents a border case.
As Schuch et al.\ \cite{Schuch08} demonstrated, it can neither guarantee an upper nor a lower bound
in general.
However, a diverging von Neumann entropy provides a lower bound \cite{Schuch08}.
More importantly, Hastings showed \cite{Hastings2007} that, if $\ket{\Psi}$ is a ground
state of a gapped local Hamiltonian, $S^{\alpha=1}$ provides an upper
bound for the truncation error of a fixed bond MPS that scales
roughly as $2^{cS_{1,j}}$, similarly to \eqref{eq:simbound}.

\section{Quantum Mutual Information}\label{sec:qmi}

The entanglement entropy thus provides a good estimate of the
simulability of states with MPS. For the truncation error of a fixed bond
MPS the relevant quantities are the entropies of subchains $S_{[1,,\ldots,j]}$.
One can, in principle, estimate these entropies during a DMRG procedure
by diagonalizing the reduced density matrices $\rho^{[1,\ldots,j]}$.
The computational effort of computing and diagonalizing
$\rho^{[1,\ldots,j]}$ will, however, grow exponentially in the length $j$ of the subchain.

An alternative is provided by the quantum mutual information (QMI)
\begin{align*}
    I_{i,j}:=S_i+S_j-S_{ij}.
\end{align*}
For two sites $i$ and $j$, $I_{i,j}$ requires only the computation of the single-site
entropies $S_i$, $S_j$ and the two-site entropy $S_{ij}$.
In \cite{Barcza2011,Rissler2006} the authors
propose minimizing the entanglement distance
\begin{align*}
    \Idist^\pm:=\pm\sum_{i,j}I_{i,j}\times|i-j|^\eta.
\end{align*}

The quantity $I_{i,j}$ measures the amount of
classical and quantum correlations between sites
$i$ and $j$. It is equal to the relative entropy
\begin{align*}
    I_{i,j}=S(\rho^{[ij]}\|\rho^{[i]}\otimes\rho^{[j]}),
\end{align*}
which is the quantum analogue of
the \emph{Kullback-Leibler divergence}.
The quantity $S(\rho^{[ij]}\|\rho^{[i]}\otimes\rho^{[j]})$
measures, in a sense, the entropic distance between the true reduced two-site
state $\rho^{[ij]}$ and its separable approximation
$\rho^{[i]}\otimes\rho^{[j]}$.
In case the sites $i$ and $j$ are uncorrelated, we have
$\rho^{[ij]}=\rho^{[i]}\otimes\rho^{[j]}$, $S_{ij}=S_i+S_j$ and
$I_{i,j}=0$. In case the correlation between $i$ and $j$
is maximal, $\rho^{[ij]}=\ket{\Psi_{ij}}\bra{\Psi_{ij}}$ is a pure state,
$S_{ij}=0$ and $I_{i,j}=S_i+S_j$.

The two-point QMI can be directly related to
the entanglement entropy of subchains as follows.
\begin{theorem}
    For any $j=1,\ldots,L-1$ and any
    $\delta=1,\ldots,\lfloor j/2\rfloor$
    \begin{align}\label{eq:main}
        S_{[1,\ldots,j]}\leq \sum_{k=1}^jS_k-\sum_{k=1}^{j-\delta}I_{k,k+\delta}.
    \end{align}
\end{theorem}
This bound holds for any entropy measure satisfying the
strong subadditivity property (SSA), see Appendix \ref{app:proof} for a proof and
Figure \ref{fig:mps} for an illustration.
\begin{figure}
    \begin{subfigure}[b]{0.4\textwidth}
        \begin{tikzpicture}
    \SetVertexStyle[MinSize=0.2\DefaultUnit,LineOpacity=0]
    \SetEdgeStyle[LineWidth=1pt]
    \Vertex[NoLabel=true,x=0, y=0, label={A}, style={shading=ball,ball color=NavyBlue}]{A}
    \Vertex[NoLabel=true,x=-1, y=-0.5, label={B}, style={shading=ball,ball color=Red}]{B}
    \Vertex[NoLabel=true,x=1, y=0.3, label={C}, style={shading=ball,ball color=NavyBlue}]{C}
    \Vertex[NoLabel=true,x=0.5, y=1, label={D}, style={shading=ball,ball color=Red}]{D}
    \Vertex[NoLabel=true,x=-1.5, y=0.8, label={E}, style={shading=ball,ball color=Red}]{E}
    \Vertex[NoLabel=true,x=-0.5, y=1.3, label={F}, style={shading=ball,ball color=Red}]{F}
    \Vertex[NoLabel=true,x=-0.5, y=-1.1, label={G}, style={shading=ball,ball color=Red}]{G}
    \Vertex[NoLabel=true,x=-0.1, y=0.3, label={H}, style={shading=ball,ball color=NavyBlue}]{H}
    \Vertex[NoLabel=true,x=-0.3, y=0.7, label={I}, style={shading=ball,ball color=NavyBlue}]{I}
    \Vertex[NoLabel=true,x=-0.8, y=0.1, label={J}, style={shading=ball,ball color=NavyBlue}]{J}
    \Vertex[NoLabel=true,x=-1.1, y=0.4, label={K}, style={shading=ball,ball color=NavyBlue}]{K}
    \Vertex[NoLabel=true,x=1.1, y=-0.4, label={L}, style={shading=ball,ball color=NavyBlue}]{L}
    \Vertex[NoLabel=true,x=0.75, y=-1.4, label={M}, style={shading=ball,ball color=Red}]{M}
    \Vertex[NoLabel=true,x=1.1, y=-0.8, label={N}, style={shading=ball,ball color=NavyBlue}]{N}
    \Vertex[NoLabel=true,x=0.1, y=-0.4, label={O}, style={shading=ball,ball color=NavyBlue}]{O}
    \Vertex[NoLabel=true,x=-0.1, y=-0.85, label={P}, style={shading=ball,ball color=NavyBlue}]{P}
    \Vertex[NoLabel=true,x=0.5, y=-0.6, label={Q}, style={shading=ball,ball color=NavyBlue}]{Q}  
    \Vertex[NoLabel=true,x=-0.3, y=-0.6, label={R}, style={shading=ball,ball color=NavyBlue}]{R}
    \Vertex[NoLabel=true,x=0.5, y=0.5, label={S}, style={shading=ball,ball color=NavyBlue}]{S}
    \Vertex[NoLabel=true,x=0.6, y=0, label={T}, style={shading=ball,ball color=NavyBlue}]{T}
    
    \Edge(E)(F)
    \Edge(E)(K)
    \Edge(K)(J)
    \Edge(J)(B)
    \Edge(B)(R)
    \Edge(R)(P)
    \Edge(P)(G)
    \Edge(G)(M)
    \Edge(M)(N)
    \Edge(N)(L)
    \Edge(L)(C)
    \Edge(C)(D)
    \Edge(D)(I)
    \Edge(S)(H)
    \Edge(S)(T)
    \Edge(T)(Q)
    \Edge(H)(A)
    \Edge(T)(O)
    \Edge(S)(I)
    \Edge(I)(F)
    
    \fill[orange,opacity=0.3] (-1.7,1.5) -- (-1.7,0) -- (-1.4,-0.2) -- (-1.2, -1)
    -- (0, -1.5) -- (1.2, -1.7) -- (1.5,0) -- (1,1) -- (0, 1.5);
    
    \node at (-0.8,0.7) {$A$};
    \node at (-2.5,2) {$\Omega$};
    
    \Vertex[NoLabel=true,x=-2, y=0, label={AA}, style={shading=ball,ball color=NavyBlue}]{AA}
    \Vertex[NoLabel=true,x=2, y=0.1, label={AB}, style={shading=ball,ball color=NavyBlue}]{AB}
    \Vertex[NoLabel=true,x=0, y=2, label={AC}, style={shading=ball,ball color=NavyBlue}]{AC}
    \Vertex[NoLabel=true,x=-0.2, y=-2, label={AD}, style={shading=ball,ball color=NavyBlue}]{AD}
    
    \Vertex[NoLabel=true,x=-2.2, y=1.2, label={AE}, style={shading=ball,ball color=NavyBlue}]{AE}
    \Vertex[NoLabel=true,x=3.2, y=1.1, label={AF}, style={shading=ball,ball color=NavyBlue}]{AF}
    \Vertex[NoLabel=true,x=1.2, y=2, label={AG}, style={shading=ball,ball color=NavyBlue}]{AG}
    \Vertex[NoLabel=true,x=2.5, y=-2, label={AH}, style={shading=ball,ball color=NavyBlue}]{AH}
    \Vertex[NoLabel=true,x=1.5, y=1.3, label={AI}, style={shading=ball,ball color=NavyBlue}]{AI}
    
    \Vertex[NoLabel=true,x=2.2, y=1.2, label={AJ},style={shading=ball,ball color=NavyBlue}]{AJ}
    \Vertex[NoLabel=true,x=2.2, y=-1.1, label={AK},style={shading=ball,ball color=NavyBlue}]{AK}
    \Vertex[NoLabel=true,x=-1.2, y=-2, label={AL},style={shading=ball,ball color=NavyBlue}]{AL}
    \Vertex[NoLabel=true,x=1.5, y=-2, label={AM},style={shading=ball,ball color=NavyBlue}]{AM}
    
    \Vertex[NoLabel=true,x=-2.5, y=-2.5, label={AN},style={shading=ball,ball color=NavyBlue}]{AN}
    \Vertex[NoLabel=true,x=-2, y=2, label={AO},style={shading=ball,ball color=NavyBlue}]{AO}
    \Vertex[NoLabel=true,x=-1, y=2.5, label={AP},style={shading=ball,ball color=NavyBlue}]{AP}
    \Vertex[NoLabel=true,x=-2.5, y=-1, label={AQ},style={shading=ball,ball color=NavyBlue}]{AQ}
    
    \Vertex[NoLabel=true,x=-2, y=-2, label={AR},style={shading=ball,ball color=NavyBlue}]{AR}
    \Vertex[NoLabel=true,x=-2.3, y=0, label={AS},style={shading=ball,ball color=NavyBlue}]{AS}
    \Vertex[NoLabel=true,x=-2.3, y=0.8, label={AT},style={shading=ball,ball color=NavyBlue}]{AT}
    \Vertex[NoLabel=true,x=2.5, y=-1, label={AU},style={shading=ball,ball color=NavyBlue}]{AU}
    
    \Vertex[NoLabel=true,x=-1.5, y=-1, label={AV},style={shading=ball,ball color=NavyBlue}]{AV}
    \Vertex[NoLabel=true,x=-3.5, y=0, label={AX},style={shading=ball,ball color=NavyBlue}]{AX}
    \Vertex[NoLabel=true,x=2.5, y=0.2, label={AY},style={shading=ball,ball color=NavyBlue}]{AY}
    \Vertex[NoLabel=true,x=0.5, y=-2.5, label={AW},style={shading=ball,ball color=NavyBlue}]{AW}
    
    \Edge(E)(AE)
    \Edge(E)(AT)
    \Edge(F)(AC)
    \Edge(B)(AV)
    \Edge(G)(AD)
    \Edge(M)(AM)
    \Edge(D)(AI)
    \Edge(AS)(AA)
    \Edge(AT)(AS)
    \Edge(AX)(AS)
    \Edge(AQ)(AS)
    \Edge(AQ)(AV)
    \Edge(AV)(AR)
    \Edge(AR)(AN)
    \Edge(AR)(AL)
    \Edge(AD)(AW)
    \Edge(AW)(AM)
    \Edge(AH)(AK)
    \Edge(AK)(AU)
    \Edge(AB)(AY)
    \Edge(AY)(AF)
    \Edge(AF)(AJ)
    \Edge(AJ)(AI)
    \Edge(AC)(AG)
    \Edge(AG)(AI)
    \Edge(AB)(AU)
    \Edge(AP)(AC)
    \Edge(AP)(AO)
\end{tikzpicture}
        \caption{Entanglement in a network.
        Subsystem $A$ is, in general, entangled with its
        environment $\Omega\setminus A$.
        The volume $|A|$ is the number of nodes in $A$,
        the area $|\partial A|$ is the number of nodes connecting to the
        environment, highlighted in red.}
    \end{subfigure}
    
    \begin{subfigure}[b]{0.4\textwidth}
        \begin{tikzpicture}
    \SetVertexStyle[MinSize=0.2\DefaultUnit,LineOpacity=0]
    \SetEdgeStyle[LineWidth=1pt]
    \Vertex[NoLabel=true,x=0, y=0, label={A}, style={shading=ball,ball color=NavyBlue}]{A}
    \Vertex[NoLabel=true,x=1, y=0, label={B}, style={shading=ball,ball color=NavyBlue}]{B}
    \Vertex[NoLabel=true,x=2, y=0, label={C}, style={shading=ball,ball color=NavyBlue}]{C}
    \Vertex[NoLabel=true,x=3, y=0, label={D}, style={shading=ball,ball color=NavyBlue}]{D}
    \Vertex[NoLabel=true,x=4, y=0, label={E}, style={shading=ball,ball color=NavyBlue}]{E}
    \Vertex[NoLabel=true,x=5, y=0, label={F}, style={shading=ball,ball color=NavyBlue}]{F}
    \Vertex[NoLabel=true,x=6, y=0, label={G}, style={shading=ball,ball color=NavyBlue}]{G}
    
    \Edge(A)(B)
    \Edge(B)(C)
    \Edge(C)(D)
    \Edge(D)(E)
    \Edge(E)(F)
    \Edge(F)(G)
    
    \draw[rounded corners, fill=orange, opacity=0.3] (-0.2, 0.2) rectangle (0.2, -0.2) {};
    
    
    \Vertex[NoLabel=true,x=0, y=-1, label={A}, style={shading=ball,ball color=NavyBlue}]{AA}
    \Vertex[NoLabel=true,x=1, y=-1, label={B}, style={shading=ball,ball color=NavyBlue}]{BB}
    \Vertex[NoLabel=true,x=2, y=-1, label={C}, style={shading=ball,ball color=NavyBlue}]{CC}
    \Vertex[NoLabel=true,x=3, y=-1, label={D}, style={shading=ball,ball color=NavyBlue}]{DD}
    \Vertex[NoLabel=true,x=4, y=-1, label={E}, style={shading=ball,ball color=NavyBlue}]{EE}
    \Vertex[NoLabel=true,x=5, y=-1, label={F}, style={shading=ball,ball color=NavyBlue}]{FF}
    \Vertex[NoLabel=true,x=6, y=-1, label={G}, style={shading=ball,ball color=NavyBlue}]{GG}
    
    \Edge(AA)(BB)
    \Edge(BB)(CC)
    \Edge(CC)(DD)
    \Edge(DD)(EE)
    \Edge(EE)(FF)
    \Edge(FF)(GG)
    
    \draw[rounded corners, fill=orange, opacity=0.3] (-0.2, -0.8) rectangle (1.2, -1.2) {};
    

    \Vertex[NoLabel=true,x=0, y=-2, label={A}, style={shading=ball,ball color=NavyBlue}]{AAA}
    \Vertex[NoLabel=true,x=1, y=-2, label={B}, style={shading=ball,ball color=NavyBlue}]{BBB}
    \Vertex[NoLabel=true,x=2, y=-2, label={C}, style={shading=ball,ball color=NavyBlue}]{CCC}
    \Vertex[NoLabel=true,x=3, y=-2, label={D}, style={shading=ball,ball color=NavyBlue}]{DDD}
    \Vertex[NoLabel=true,x=4, y=-2, label={E}, style={shading=ball,ball color=NavyBlue}]{EEE}
    \Vertex[NoLabel=true,x=5, y=-2, label={F}, style={shading=ball,ball color=NavyBlue}]{FFF}
    \Vertex[NoLabel=true,x=6, y=-2, label={G}, style={shading=ball,ball color=NavyBlue}]{GGG}
    
    \Edge(AAA)(BBB)
    \Edge(BBB)(CCC)
    \Edge(CCC)(DDD)
    \Edge(DDD)(EEE)
    \Edge(EEE)(FFF)
    \Edge(FFF)(GGG)
    
    \draw[rounded corners, fill=orange, opacity=0.3] (-0.2, -1.8) rectangle (4.2, -2.2) {};
    
    \node (1) at (3, -2.5) {};
    \node (2) at (3, -3.5) {};
    \draw[->] (1) -- (2);


    \Vertex[NoLabel=true,x=0, y=-4, label={A}, style={shading=ball,ball color=NavyBlue}]{AAAA}
    \Vertex[NoLabel=true,x=1, y=-4, label={B}, style={shading=ball,ball color=NavyBlue}]{BBBB}
    \Vertex[NoLabel=true,x=2, y=-4, label={C}, style={shading=ball,ball color=NavyBlue}]{CCCC}
    \Vertex[NoLabel=true,x=3, y=-4, label={D}, style={shading=ball,ball color=NavyBlue}]{DDDD}
    \Vertex[NoLabel=true,x=4, y=-4, label={E}, style={shading=ball,ball color=NavyBlue}]{EEEE}
    \Vertex[NoLabel=true,x=5, y=-4, label={F}, style={shading=ball,ball color=NavyBlue}]{FFFF}
    \Vertex[NoLabel=true,x=6, y=-4, label={G}, style={shading=ball,ball color=NavyBlue}]{GGGG}
    
    \Edge(AAAA)(BBBB)
    \Edge(BBBB)(CCCC)
    \Edge(CCCC)(DDDD)
    \Edge(DDDD)(EEEE)
    \Edge(EEEE)(FFFF)
    \Edge(FFFF)(GGGG)
    
    \node at (0.5,-4.5) {$-$};
    \node at (1.5,-4.5) {$-$};
    \node at (2.5,-4.5) {$-$};
    \node at (3.5,-4.5) {$-$};
    
    \Edge[Direct,bend=-45](AAAA)(BBBB)
    \Edge[Direct,bend=-45](BBBB)(CCCC)
    \Edge[Direct,bend=-45](CCCC)(DDDD)
    \Edge[Direct,bend=-45](DDDD)(EEEE)

    \draw[rounded corners, fill=orange, opacity=0.3] (-0.2, -3.8) rectangle (0.2, -4.2) {};
    \draw[rounded corners, fill=orange, opacity=0.3] (0.8, -3.8) rectangle (1.2, -4.2) {};
    \draw[rounded corners, fill=orange, opacity=0.3] (1.8, -3.8) rectangle (2.2, -4.2) {};
    \draw[rounded corners, fill=orange, opacity=0.3] (2.8, -3.8) rectangle (3.2, -4.2) {};
    \draw[rounded corners, fill=orange, opacity=0.3] (3.8, -3.8) rectangle (4.2, -4.2) {};


    \Vertex[NoLabel=true,x=0, y=-5, label={A}, style={shading=ball,ball color=NavyBlue}]{AAAAA}
    \Vertex[NoLabel=true,x=1, y=-5, label={B}, style={shading=ball,ball color=NavyBlue}]{BBBBB}
    \Vertex[NoLabel=true,x=2, y=-5, label={C}, style={shading=ball,ball color=NavyBlue}]{CCCCC}
    \Vertex[NoLabel=true,x=3, y=-5, label={D}, style={shading=ball,ball color=NavyBlue}]{DDDDD}
    \Vertex[NoLabel=true,x=4, y=-5, label={E}, style={shading=ball,ball color=NavyBlue}]{EEEEE}
    \Vertex[NoLabel=true,x=5, y=-5, label={F}, style={shading=ball,ball color=NavyBlue}]{FFFFF}
    \Vertex[NoLabel=true,x=6, y=-5, label={G}, style={shading=ball,ball color=NavyBlue}]{GGGGG}
    
    \Edge(AAAAA)(BBBBB)
    \Edge(BBBBB)(CCCCC)
    \Edge(CCCCC)(DDDDD)
    \Edge(DDDDD)(EEEEE)
    \Edge(EEEEE)(FFFFF)
    \Edge(FFFFF)(GGGGG)
    
    \Edge[Direct,bend=-45](AAAAA)(CCCCC)
    \Edge[Direct,bend=-45](BBBBB)(DDDDD)
    \Edge[Direct,bend=-45](CCCCC)(EEEEE)
    
    \node at (1,-5.75) {$-$};
    \node at (2,-5.75) {$-$};
    \node at (3,-5.75) {$-$};

    \draw[rounded corners, fill=orange, opacity=0.3] (-0.2, -4.8) rectangle (0.2, -5.2) {};
    \draw[rounded corners, fill=orange, opacity=0.3] (0.8, -4.8) rectangle (1.2, -5.2) {};
    \draw[rounded corners, fill=orange, opacity=0.3] (1.8, -4.8) rectangle (2.2, -5.2) {};
    \draw[rounded corners, fill=orange, opacity=0.3] (2.8, -4.8) rectangle (3.2, -5.2) {};
    \draw[rounded corners, fill=orange, opacity=0.3] (3.8, -4.8) rectangle (4.2, -5.2) {};

\end{tikzpicture}
        \caption{For an MPS the entanglement entropy of subchains
        $\set{1,\ldots,j}$ controls simulability. As in \eqref{eq:main},
        the subchain entropy can be broken up into the total correlation
        minus QMI corrections.}
        \label{fig:mps}    
    \end{subfigure}
    \caption{}
    \label{fig:tensors}
\end{figure}
\section{Discussion}\label{sec:discuss}

From \eqref{eq:main} and Figure \ref{fig:mps}, we see that we can sum up
$I_{i,j}$ in different ways, and one can thus ask which bound is sharper.
For
further reference, note that the von Neumann entropy satisfies
the following subadditivity (SA) and strong subadditivity (SSA)
properties
\begin{align}
    S(\rho^{[AB]})&\leq S(\rho^{[A]})+S(\rho^{[B]}),\label{sa}\tag{SA}\\
    S(\rho^{[AB]})+S(\rho^{[BC]})&\geq S(\rho^{[B]})+S(\rho^{[ABC]}).\label{ssa}\tag{SSA}
\end{align}

Consider first $\delta=1$. 
In \eqref{eq:main},
one ``chains'' the entropy together via a series of inequalities
as
\begin{align}\label{eq:chains}
    S(\rho^{[1,\ldots,k]})+S(\rho^{[k,k+1]})\geq
    S(\rho^{[k]})+S(\rho^{[1,\ldots,k+1]}),
\end{align}
using \eqref{ssa}.
Denoting by $I(A:B)$ the QMI between subsystems
$A$ and $B$, we can write
\begin{align*}
    S(\rho^{[1,\ldots,k+1]})&+S(\rho^{[k]})\\
    &+[I([1,\ldots,k]:k+1)-I(k:k+1)]\\
    &=S(\rho^{[1,\ldots,k]})+S(\rho^{[k,k+1]}).
\end{align*}
The difference between the two sides of inequality \eqref{eq:chains}
is the amount of entanglement the site
$k+1$ shares with the site $k$ vs.\ the subchain $\set{1,\ldots,k}$.
This difference is bounded by the single-site entropy.
Using the symmetry
$S(\rho^{[1,\ldots,k]})=S(\rho^{[k+1,\ldots,L]})$
and \eqref{sa}, we namely have
\begin{align*}
    I([1,\ldots,k]:k+1)-I(k:k+1)\leq 2S(\rho^{[k+1]}).
\end{align*}
The single-site entropy is itself bounded by the size
of the local Hilbert space. I.e.,
in \eqref{eq:chains} we overestimate by at most $2\log_2(d)$.
However, overall we need to ``chain'' such inequalities $j-1$ times.
In total for \eqref{eq:main}, in the worst case, one overestimates by
$2(j-2)\log_2(d)$.

Consider now instead $\delta=2$. Here one uses both \eqref{ssa} in the form
of
\begin{align}\label{eq:usessa}
    S(\rho^{[1,3,\ldots,k]})+S(\rho^{[k,k+2]})\geq S(\rho^{[k]})+
    S(\rho^{[1,3,\ldots,k+2]}),
\end{align}
as well as \eqref{sa}
\begin{align}\label{eq:usesa}
    S(\rho^{[1,3,\ldots,j-1]})+S(\rho^{[2,4,\ldots,j]})\geq
    S(\rho^{[1,2,3,\ldots,j]}).
\end{align}
In \eqref{eq:usessa} we once again overestimate by at most
$2S(\rho^{[k+2]})\leq 2\log_2(d)$.
However, the number of such chained inequalities is now
reduced to roughly half $j/2$.
On the other hand, we apply \eqref{eq:usesa} once and
we overestimate by
$I([1,3,\ldots,j-1]:[2,4,\ldots,j])$ which, in the worst case,
is of the order $(j/2)\log_2(d)$. Thus, in total we once again overestimate
by at most $2(j-2)\log_2(d)$.

The same reasoning applies to any $\delta$ with the overestimation
always being $2(j-2)\log_2(d)$.
The question which upper
bound provides a sharper estimate is then related to comparing
two-point QMI $I_{i,j}$ to higher order QMI
such as $I([1,3,5,\ldots,j-1]:[2,4,\ldots,j])$ and similarly
for different $\delta$. In other words, we cannot reliably estimate this
information by using two-point QMI
$I_{i,j}$ only.

Using two-point QMI can be seen as a ``first order'' approximation to the block
entropy. Specifically, estimating block entropy by the \emph{total} entropy
$\sum_kS_k$, we overestimate in the order of $\mathrm{vol}-1$.
Correcting total entropy with two-point QMI as in \eqref{eq:main}
we overestimate in the order of $\mathrm{vol}-2$.
Using higher-order corrections with correlations between blocks with up to 3 sites,
we overestimate by $\mathrm{vol}-3$ and so on.

One can exploit the information provided by the bounds for different
$\delta$ by combining them into a weighted
average, e.g., for
$1/c_j=\sum_{\delta=1}^{\lfloor j/2\rfloor}\delta^{-2}$
\begin{align}\label{eq:Ij}
    \hat{I}_j:=\sum_{k=1}^jS_k-c_j\sum_{\delta=1}^{\lfloor j/2\rfloor}
    \delta^{-2}\sum_{k=1}^{j-\delta}I_{k,k+\delta}
\end{align}
This corresponds conceptually to $\Idist^-$ of \cite{Rissler2006}
from \eqref{eq:Idist} with $\eta=-2$.

Combining \eqref{eq:Ij} with the discussion in Section \ref{sec:ent}
and \eqref{eq:simbound} suggests
that an appropriate cost function to minimize for control of the total
truncation error and control of the entropy of the MPS, is
\begin{align}\label{eq:Imps}
    \hat{I}_{\mathrm{MPS}}=\log_2\sum_{j=1}^{L-1}2^{\hat{I}_j}.
\end{align} 
The above quantity is the $\mathrm{LogSumExp}$ function and can be approximated via
\begin{align*}
    \check{I}_{\mathrm{MPS}}=\max_{j=1,\ldots,L-1}\hat{I}_j,
\end{align*}
with $\check{I}_{\mathrm{MPS}}<\hat{I}_{\mathrm{MPS}}\leq\check{I}_{\mathrm{MPS}}+\log_2(L-1)$.

\subsection{Other Entropy Measures}

For a more rigorous error control, one could consider the R\'{e}nyi entropy
$S^\alpha$ for $0<\alpha<1$ instead. However, R\'{e}nyi
satisfies only a weak subadditivity property (WSA)
\begin{align*}
    S^\alpha(\rho^{[A]})-S^0(\rho^{[B]})\leq
    S^\alpha(\rho^{[AB]})\leq
    S^\alpha(\rho^{[A]})+S^0(\rho^{[B]}).
\end{align*}
In particular, QMI $I_{i,j}$ defined for
$S^\alpha$ with $\alpha\neq 1$ is not guaranteed to be positive and
does not necessarily make sense as a measure of correlation.
The WSA can still be used to estimate $S^\alpha_{[1,\ldots,j]}$ by chaining
differences of two-point entanglement entropies. However, the resulting upper
bound is too crude to be useful.

One can also replace $I_{i,j}$ with an analogous quantity for the R\'{e}nyi
divergence.
The \emph{sandwiched quantum R\'{e}nyi relative
entropy} satisfies a version of the SSA as shown in \cite{Muller2013}.
This could provide a tighter control of the subchain entropy
$S^\alpha_{[1,\ldots,j]}$, though we have not analyzed this further.

Furthermore, R\'{e}nyi was numerically shown \cite{Camilo2019}
to satisfy SSA for
$0\leq \alpha\leq 2$ in the case of Gaussian bosons and
$0\leq\alpha\leq 1.3$ in the case of Gaussian fermions.
This suggests that the same orbital optimization as above
with $S^\alpha$ replacing the von Neumann entropy may still
provide good results for important classes of physical states.
However, overall, we expect the von Neumann entropy will provide
better control of the total DMRG error and information loss,
albeit not with the same generality as the
R\'{e}nyi entropy.

\begin{acknowledgements}
    I would like to thank Alexander Nüßeler and Anthony Nouy for helpful remarks on this work.
\end{acknowledgements}

\appendix

\section{Proof of \eqref{eq:main}}\label{app:proof}

For any $\delta=1,\ldots,\lfloor j/2\rfloor$, we can write
\begin{align}\label{eq:aux}
    &\sum_{k=1}^\delta S(\rho^{[k]})+
    \sum_{k=j-\delta+1}^jS(\rho^{[k]})
    +2\sum_{k=\delta+1}^{j-\delta}S(\rho^{[k]})\\
    &-
    \sum_{k=1}^{j-\delta}I(k:k+\delta)\notag\\
    &=\sum_{k=1}^{j-\delta}S(\rho^{[k]})+S(\rho^{[k+\delta]})-I(k:k+\delta)\notag\\
    &=\sum_{k=1}^{j-\delta}S(\rho^{[k,k+\delta]})\notag
\end{align}
We chain together the terms with a common site
using SSA
\begin{align*}
    \sum_{k=1}^{j-\delta}S(\rho^{[k,k+\delta]})
    &\geq \sum_{k=1}^\delta
    S(\rho^{[k,k+\delta, k+2\delta,\ldots,k+\lfloor(j-k)/\delta\rfloor\delta]})\\
    &+\sum_{k=\delta+1}^{j-\delta}
    S(\rho^{[k]})
\end{align*}
We chain together the disjoint many-site entropies using SA
\begin{align*}
    \sum_{k=1}^\delta
    S(\rho^{[k,k+\delta, k+2\delta,\ldots,k+\lfloor(j-k)/\delta\rfloor\delta]})
    \geq
    S(\rho^{[1,\ldots,j]})
\end{align*}
The single-site entropies cancel with the
single-site entropies in \eqref{eq:aux}. And thus we obtain
\begin{align*}
    S(\rho^{[1,\ldots,j]})\leq\sum_{k=1}^j S_k
    -\sum_{k=1}^{j-\delta}I(k:k+\delta).
\end{align*}

\section{Criteria of \cite{Dupuy21}}\label{app:misong}
Dupuy {\&} Friesecke \cite{Dupuy21} observed that, if $\ket{\Psi}$ is
the second quantization of a Slater determinant,
e.g., it is a non-interacting fermionic state, then
the singular values of a representation in any a priori fixed
single particle basis obey the relation
\begin{align}\label{eq:symm}
    \sigma_k^{[1,\ldots,j]}\sigma_{M-k}^{[1,\ldots,j]}=
    p_j,
\end{align}
where $\sigma_M^{[1,\ldots,j]}$ is the last nonzero
singular value, and
the factor $p_j$ is a constant depending only on the subchain $\set{1,\ldots,j}$.
The authors \cite{Dupuy21} then suggest optimizing the orbital structure
by minimizing $p_j$. This scheme is tested on upto 3-term FCI expansions
and outperforms the orbital ordering scheme of \cite{Barcza2011} by several orders of
magnitude.

Is there a relationship to optimizing the entanglement distance
$\Idist^\pm$? The symmetry in \eqref{eq:symm}
provides implicitly a restriction on the entropy values
$S_{[1,\ldots,j]}$ can attain. This by itself, however, is not sufficient
to justify small entropy or, equivalently, fast decay of singular
values. It is not difficult to see that one can construct an entropy
maximizing sequence, obeying \eqref{eq:symm}, that will
have entropy roughly in the order of $\log_2(M/2)$, i.e.,
this restriction reduces the maximal entropy content by at most $1$.

In that sense, $\Idist^\pm$ from \eqref{eq:Idist}
or $\hat{I}_{\mathrm{MPS}}$ from \eqref{eq:Imps}
provides a more rigorous
control of the truncation error and entropy due to \eqref{eq:main}.
However, using \eqref{eq:symm} (and modifications provided
in \cite{Dupuy21}) as a criteria may still perform better in practice.

\section{Tree Tensor Networks}

\begin{figure}
    \begin{tikzpicture}
    \SetVertexStyle[MinSize=0.2\DefaultUnit,LineOpacity=0]
    \SetEdgeStyle[LineWidth=1pt]
    \Vertex[NoLabel=true,x=0, y=0, label={A}, style={shading=ball,ball color=NavyBlue}]{A}
    \Vertex[NoLabel=true,x=1, y=0, label={B}, style={shading=ball,ball color=NavyBlue}]{B}
    \Vertex[NoLabel=true,x=2, y=0, label={C}, style={shading=ball,ball color=NavyBlue}]{C}
    \Vertex[NoLabel=true,x=3, y=0, label={D}, style={shading=ball,ball color=NavyBlue}]{D}
    \Vertex[NoLabel=true,x=4, y=0, label={E}, style={shading=ball,ball color=NavyBlue}]{E}
    \Vertex[NoLabel=true,x=5, y=0, label={F}, style={shading=ball,ball color=NavyBlue}]{F}
    \Vertex[NoLabel=true,x=6, y=0, label={G}, style={shading=ball,ball color=NavyBlue}]{G}
    \Vertex[NoLabel=true,x=7, y=0, label={A}, style={shading=ball,ball color=NavyBlue}]{H}
    
    \Vertex[NoLabel=true,x=0.5, y=1, label={E}, style={shading=ball,ball color=NavyBlue}]{AA}
    \Vertex[NoLabel=true,x=2.5, y=1, label={F}, style={shading=ball,ball color=NavyBlue}]{BB}
    \Vertex[NoLabel=true,x=4.5, y=1, label={G}, style={shading=ball,ball color=NavyBlue}]{CC}
    \Vertex[NoLabel=true,x=6.5, y=1, label={A}, style={shading=ball,ball color=NavyBlue}]{DD}
    
    \Vertex[NoLabel=true,x=1.5, y=2, label={G}, style={shading=ball,ball color=NavyBlue}]{AAA}
    \Vertex[NoLabel=true,x=5.5, y=2, label={A}, style={shading=ball,ball color=NavyBlue}]{BBB}
    \Vertex[NoLabel=true,x=3.5, y=3, label={A}, style={shading=ball,ball color=NavyBlue}]{root}
    
    \Edge(AA)(A)
    \Edge(AA)(B)
    \Edge(BB)(C)
    \Edge(BB)(D)
    \Edge(CC)(E)
    \Edge(CC)(F)
    \Edge(DD)(G)
    \Edge(DD)(H)
    
    \Edge(AAA)(AA)
    \Edge(AAA)(BB)
    \Edge(BBB)(CC)
    \Edge(BBB)(DD)
    
    \Edge(root)(AAA)
    \Edge(root)(BBB)
    
    \node (1) at (0,-1) {};
    \node (2) at (7,-1) {};
    \node (3) at (-0.4,0) {};
    \node (4) at (-0.4,3) {};
    \draw[<->] (1) -- (2) node[midway, below] {$L$};
    \draw[<->] (3) -- (4) node[midway, rotate=90, above] {$\log_2(L)$};
\end{tikzpicture}
    \caption{Tree Tensor Network.}
    \label{fig:ttn}
\end{figure}
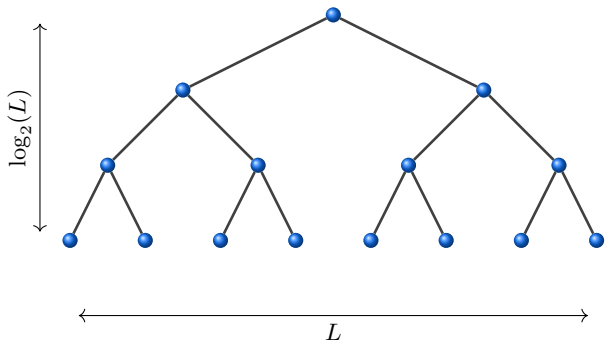

We can replace an MPS with a tree tensor network and repeat the
above considerations. E.g., consider a perfectly balanced binary tree
as in Figure \ref{fig:ttn}. An idealized optimization scheme would order the orbitals in each
layer with $\log_2(L)$ layers in total. Using only two-point QMI, however,
does not provide enough information to reliably optimize the entire
tree topology. In principle, one can use the same ordering optimization
as for MPS -- ordering orbitals on the lowest level only
-- as an approximation.

Alternatively, we can perform a finer analysis of the relevant entanglement entropies
in Figure \ref{fig:ttn}, using only two-point QMI.
Similarly to \eqref{eq:Ij}, define
for $1/c_{ij}=\sum_{\delta=1}^{\lfloor (j-i+1)/2\rfloor}\delta^{-2}$
\begin{align*}
    \hat{I}_{i,j}:=\sum_{k=i}^jS_k-\sum_{\delta=1}^{\lfloor (j-i+1)/2\rfloor}
    c_{i,j}\delta^{-2}
    \sum_{k=i}^{j-\delta}I_{k,k+\delta}.
\end{align*}
Then, for a perfectly balanced binary tree with $L=2^M$, set
\begin{align*}
    \hat{I}_{\mathrm{tree}}=\log_2
    \sum_{l=1}^{M-1}\sum_{i=1}^{2^{M-l}}2^{\hat{I}_{2^li-1,2^li}},
\end{align*}
as a measure of entropy or approximability of the tree.

\section{Exact vs.\ Approximate Low-Rank}\label{sec:apphart}

In multilinear algebra, an important concept in the
theory of tensor networks are so-called minimal subspaces \cite{Falco2012}:
for a subsystem $A\subset\Omega$,
a pure state $\ket{\Psi}$ on $\Omega$
and a corresponding
reduced density matrix $\rho^{[A]}=\tr_{\Omega\setminus A}(\ket{\Psi}\bra{\Psi})$
\begin{align*}
    \Umin^{[A]}:=\range(\rho^{[A]}).
\end{align*}
A well-known \emph{hierarchy} property states that
for any disjoint subsystems $A,B\subset\Omega$, one has
\begin{align*}
    \Umin^{[AB]}\subset\Umin^{[A]}\otimes\Umin^{[B]}.
\end{align*}
This is also a well-known property in the theory
of renormilation groups (RG) -- after
an RG transformation to
the coarser scale $AB$ (or simply \emph{coarse graining}), the corresponding
coarse grained Hilbert space is contained in the tensor product of the
Hilbert spaces on the finer scales.

One consequence of this hierarchy is the relationship between the corresponding
\emph{exact} bond dimensions (or \emph{ranks})
of $\ket{\Psi}$
\begin{align}\label{eq:ranks}
    \chi^{[AB]}\leq\chi^{[A]}\chi^{[B]}.
\end{align}
The ratio of the two quantities above
is sometimes taken as a measure of the efficiency
of an exact tensor network representation. Note that
\eqref{eq:ranks} is equivalent to the SA property of the Hartley
entropy
\begin{align*}
    S^0(\rho^{[AB]})\leq S^0(\rho^{[A]})+S^0(\rho^{[B]}).
\end{align*}
For $\alpha=0$, $S^\alpha$ quantifies ``low-rankness'' for
exact low-rank representations, while $\alpha>0$ provides
a finer grading of low-rank approximability.

\bibliography{references}

\end{document}